# Quantifying angular momentum of coherently driven circular phonons

R. Mankowsky[1*], S. Zerdane[1], S.-W. Huang[1], M. Sander[1], X. Liu[1], D. Babich[1], Martina Basini[2], Puneet Kaur[3], Jan-Chi Yang[3,4], Michael Fechner[5], Urs Staub[1], H.T. Lemke[1]

[*]Corresponding author. Email: roman.mankowsky@psi.ch
[1]*Paul Scherrer Institute, 5232 Villigen PSI, Switzerland*
[2]*Institute for Quantum Electronics, Physics Department, ETH Zürich, Zürich, Switzerland*
[3]*Department of Physics, National Cheng Kung University, Tainan, Taiwan*
[4]*Center for Quantum Frontiers of Research & Technology (QFort), National Cheng Kung University, Tainan, Taiwan*
[5]*Max Planck Institute for the Structure and Dynamics of Matter, Hamburg, Germany*

**Abstract**
The use of intense terahertz (THz) pulses to manipulate low-energy excitations offers a powerful approach for ultrafast control of electronic and magnetic properties in materials. Theory suggests that circular ionic motions driven by THz fields carry angular momentum, potentially generating internal magnetic fields. Recent experiments in nonmagnetic $SrTiO_3$ (STO) have hinted at such THz-induced fields, but their origin remains debated. Here, we employ ultrafast x-ray diffraction to resolve the time-dependent ionic trajectories in STO following excitation by circularly polarized THz pulses. Our analysis reveals that oxygen ions, despite their lower mass, contribute ~90% of the phonon angular momentum. The resulting imbalance between the negatively and positively charged ions provides a clear explanation for the mechanism behind induced magnetism in STO. This work further provides the first quantitative measurement of circular ionic motions and their angular momentum and establishes a general methodology for the investigation of angular momentum transfer in solids, paving the way for new strategies to control topological phonon transport and phonon-driven magnetism in quantum materials.

**Introduction:**
The connection between magnetization and angular momentum in ferromagnetic materials, first demonstrated by the Einstein-de-Haas (EdH) effect in 1915[1,2], has regained prominence with the advent of ultrafast laser techniques. Ultrafast demagnetization experiments[3,4] required rapid angular momentum transfer from spins to the lattice, sparking studies on the ultrafast EdH effect[5,6] and very recently, its inverse effect. Theoretically proposed in 2018[7], magnetophononics posits that circular ionic motions driven by terahertz fields carry angular momentum and can induce magnetization. Recent experiments report magnetic fields exceeding 10 T and even ferromagnetic switching[8–12], yet these findings starkly contrast with first-principles predictions, which estimate effects orders of magnitude weaker[13–16]. The angular momentum of chiral phonons[17] has been explored in chiral materials ranging from molecules[18] to condensed matter[19], notably through the phonon-Zeeman effect[20], circular polarized Raman[21] and inelastic x-ray scattering[22], and the observation of angular momentum transfer between phonons has recently been reported[23]. However, direct quantification of atomic trajectories and angular momentum during and after the excitation has remained elusive.

**Main text:**
Here, we use ultrafast x-ray diffraction to resolve ionic trajectories in $SrTiO_3$ (STO), providing the first time-resolved measurement of phonon angular momentum and establishing a general methodology for its investigation in quantum materials. We obtain the circular trajectories of all ions in the unit cell in real time and extract the angular momentum carried by them. Due to the larger radius of their circular motion, the oxygen ions account for 90% of the phonon angular momentum despite their small mass compared to the titanium and strontium ions. Our measurements provide a general methodology that can be routinely applied to most materials to quantify the trajectory of atomic motions.

Upon cooling, STO shows strong softening of an infrared active phonon mode, the polar soft mode, indicating a ferroelectric instability[24,25]. However, instead of forming ferroelectric long-range order, STO attains a quantum paraelectric ground state with a strong instability to strain, pressure and electric fields, which has triggered much interest in its phononic properties[26,27]. Recent time-resolved femtosecond x-ray diffraction studies have demonstrated strong coupling between the polar mode and transverse acoustic modes[28] as well as phonon upconversion[29], two effects, which are both related to the large anharmonicity of its energy potential. Optical experiments found signatures of emergent ferroelectric order induced by resonant phonon excitation[30–32] as well as signatures of magnetic fields induced by circular THz excitation of the polar mode in the paraelectric state[9]. In all these cases, the induced atomic motions represent the basic origin for the observations, highlighting the importance for quantifying their time-dependent trajectories and amplitudes.

We investigated a free-standing STO film on a Si wafer with 80 nm thickness, well below the THz penetration depth $d = 300\ nm$ at the polar mode resonance frequency. This ensures a homogeneous excitation of the probed volume. Static XRD and optical measurements reveal a transition from a paraelectric to a ferroelectric state below $T_c \approx 82\ K$, consistent with observations in STO films subjected to small residual strain fields[25,33]. All measurements were done in the paraelectric cubic phase at 320 K, where the polar mode frequency is 2.9 THz. The sample was excited with broadband single-cycle THz pulses generated by optical rectification in a DSTMS crystal (main frequency content: 1.5-4 THz). The pulse polarization was converted from linear to circular using an 89µm thick Y-cut quartz crystal, which acts as a $\lambda/4$ waveplate at the polar mode frequency. The incident THz pulses had peak electric fields of 400 kV cm$^{-1}$. After accounting for reflection at the interface, the field inside the film $E_{film}$ is reduced to 180 kV cm⁻¹, which remains below the threshold for inducing anharmonic phenomena such as phonon up-conversion[29,34]. We refer to Methods sections 1 and 2 and Extended Data Figures 1-5 for details about the experimental setup and THz electric field characterization.

The experiment was performed at the Bernina beamline of the SwissFEL Free Electron Laser [35,36], which provided 50 fs FWHM x-ray pulses with a photon energy of 8.3keV, defined with a Si (111) double crystal monochromator. The experimental geometry is shown in Fig. 1. The THz pulses (orange) impinge on the sample at an angle of 75° with respect to the sample surface plane, collinear with the x-rays (yellow). Owing to the large refractive index of silicon at 2.9 THz ($n = 3.14$), the THz pulses are refracted such that they propagate predominantly along the [001] surface normal into the sample, with their electric field polarized within the surface plane. In order to quantitatively measure the polarization state and amplitude of the induced ionic motions, the sample was oriented to simultaneously record the structural diffraction peaks with miller indices (hkl) = (203) and (023) on a Jungfrau detector[37]. The intensity of these peaks is modulated by atomic motions along the orthogonal $x$ and $y$ directions ([100], [010]), respectively. Therefore, the in-plane ionic motions can be fully reconstructed from the changes in diffracted intensity recorded in a single time scan.

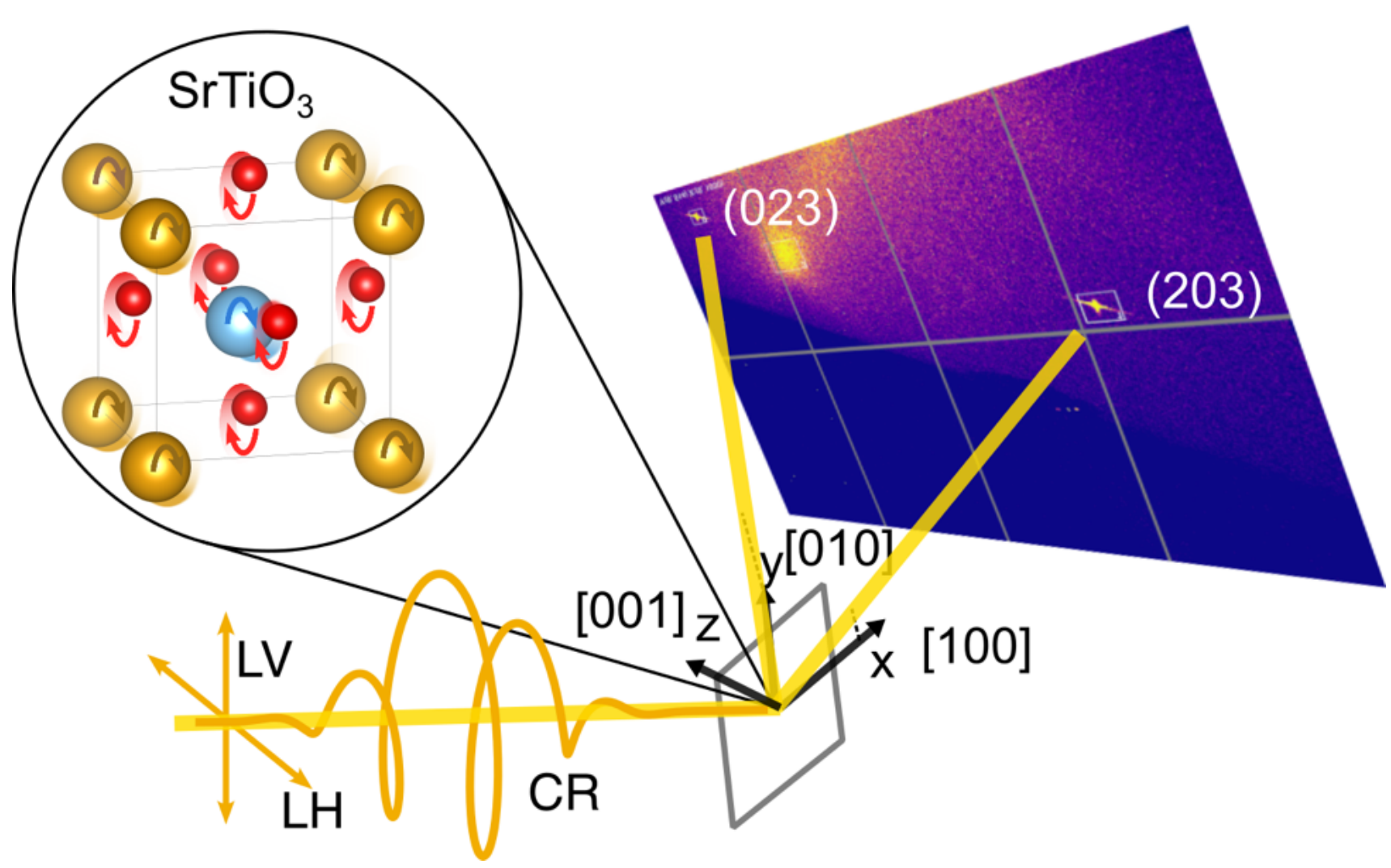

**Fig. 1: Experimental setup.** High-field THz pulses with linear vertical (LV), horizontal (LH) as well as circular left (CL) and right polarization (CR) were used to excite the STO thin film with [001] surface normal. The induced in-plane ionic motions within the unit cell modulate the diffracted intensity of the (203) and (023) Bragg peaks, which are measured simultaneously.

The results for varying THz polarization are summarized in Fig. 2, showing the diffracted intensity changes of the (203) peak in blue and the (023) peak in green. For linear vertical (LV) THz polarization, the electric field is polarized along the [110] in-plane direction with equal projections onto the [100] and [010] directions. Consequently, both diffraction peaks exhibit the same response. In contrast, only the (023) peak response flips sign (opposite phase) when the polarization is changed to linear horizontal (LH), parallel to the [$1\bar{1}0$] direction. Circular left (CL) and circular right (CR) polarizations are defined with respect to the sample's surface normal, i.e., from the perspective of an observer looking along the surface normal and opposite to the THz propagation direction. Under circular THz excitation, the induced intensity changes of the two diffraction peaks show a relative delay of $\pm 85$ fs at the maximal response, corresponding to a phase difference of $\pm\pi/2$ at 2.9 THz, respectively.

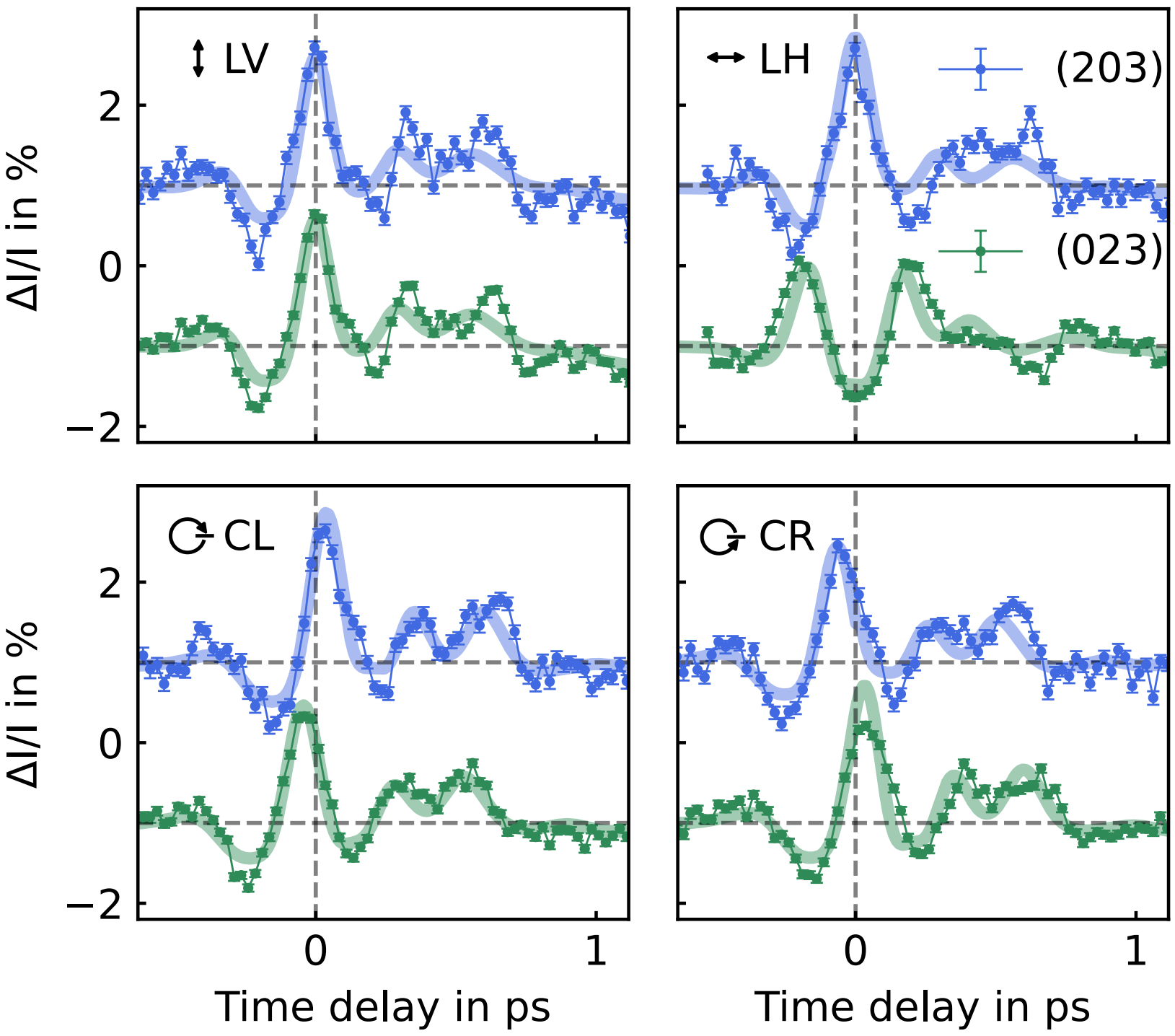

**Fig. 2: THz polarization dependence.** The relative change in diffraction intensity of the (203) and (023) diffraction peaks are shown in blue and green and offset by $\pm 1\%$, respectively. The two traces show a phase difference in their response of 0 and $\pi$ for linear vertical (LV) and linear horizontal (LH), and $\pm\pi/2$ for circular left (CL) and right (CR) THz polarization. The solid blue and green lines show the calculated change in diffraction intensity obtained from a simulation of the atomic motions using the THz field measured by electrooptic sampling. Error bars reflect the 1σ (67%) confidence interval.

In the following, we quantify the atomic displacements $\Delta\boldsymbol{r}(t) = Q_x(t)\cdot\boldsymbol{\varepsilon}_x + Q_y(t)\cdot\boldsymbol{\varepsilon}_y$ associated with observed signals, where $n = x, y$ denote the two in-plane directions, $Q_n$ is the amplitude of mode $n$, and $\boldsymbol{\varepsilon}_n$ is the corresponding displacement eigenvector describing the relative atomic motions. For a polar mode, the collective ionic displacement is directly linked to the macroscopic polarization induced by the THz electric field. We therefore determine the mode amplitudes from the time-dependent polarization $P_n(t)$ via

$$Q_n(t) = \frac{P_n(t)}{Z^*},$$

where $Z^*$ is the effective charge associated with the mode. Both $Q_n$ and $Z^*$ are mass-normalized, with units of Å$\sqrt{\mathrm{u}}$ and $1/\sqrt{\mathrm{u}}$, respectively. The displacement eigenvectors $\boldsymbol{\varepsilon}_n$ (units $1/\sqrt{\mathrm{u}}$) and their mass-normalized counterparts $\tilde{\boldsymbol{\varepsilon}}_n$, obtained from density functional theory, are listed in Extended Data Tables 1-3. More details are given in Methods section 3. The frequency-dependent polarization per unit cell is modelled as

$$P_n(\omega) = V_{uc}\varepsilon_0 S \cdot \frac{\omega_0^2}{\omega_0^2 - \omega^2 - i\Gamma\omega} E_n(\omega),$$

where $V_{uc}$ is the unit cell volume, $S = 1100$ the oscillator strength, and $\Gamma = 1.8$ THz the damping, as determined from time-domain spectroscopy on a 40nm $SrTiO_3$ film (see Supplementary Information section 1). The resonance frequency was shifted from $\omega_0 =$ 3.3 to 2.9 THz to match the resonance frequency of the measured 80nm film. The THz electric field $E_n(t)$ is obtained by electrooptic sampling[38] and corrected for reflection losses due to electronic screening. The time-dependent polarization $P_n(t)$ is then recovered by inverse Fourier transform.

With given mode amplitudes $Q_n(t)$, the position of atom $j$ evolves as $\mathbf{r}_j(t) = \mathbf{r}_{j0} + Q_x(t)\,\boldsymbol{\varepsilon}_{jx} + Q_y(t)\,\boldsymbol{\varepsilon}_{jy}$, where $\mathbf{r}_{j0}$ denotes the equilibrium position. The resulting modulation of the (203) and (023) diffraction peak intensities is calculated from the time-dependent structure factor

$$F_{hkl}(t) = \sum_j f_j \; e^{-i\mathbf{G}_{hkl}\cdot\mathbf{r}_j(t)},$$

where $f_j$ is the scattering factor of atom $j$ and $\mathbf{G}_{hkl}$ the reciprocal lattice vector. The normalized diffracted intensity is then

$$\frac{I(t)}{I_0} = \frac{|\,F_{hkl}(t)\,|^2}{|\,F_{hkl}(0)\,|^2} + g(t),$$

where $g(t)$ accounts for slowly varying strain-induced background contributions due to slow shifts of the diffraction peak positions along the out-of plane direction (see Methods Section 3 and Extended Data Figures 6 and 7). Calculations of the diffracted intensity for varying amplitude of the soft mode are displayed in Extended Data Figure 6. The solid lines in Fig. 2 show fits obtained with this approach, which accurately reproduce both the amplitude and the relative phase of the measured responses for different THz polarizations.

The only parameter adjusted to the diffraction data is its effective charge, which linearly modulates the amplitude of the atomic motions, $Z^* = 15.4\ \mathrm{e}/\sqrt{\mathrm{u}}$. We note that $Z^*$ can also be evaluated from the oscillator parameters, which gives a smaller value of $4.0\ \mathrm{e}/\sqrt{\mathrm{u}}$. This discrepancy could be related to an enhancement of the THz field inside the film due to multiple reflections at the surface and the STO/Si interfaces. Evaluating the Fabry-Pérot factor gives a value of around 1.9 at 3 THz, accounting for part of the discrepancy.

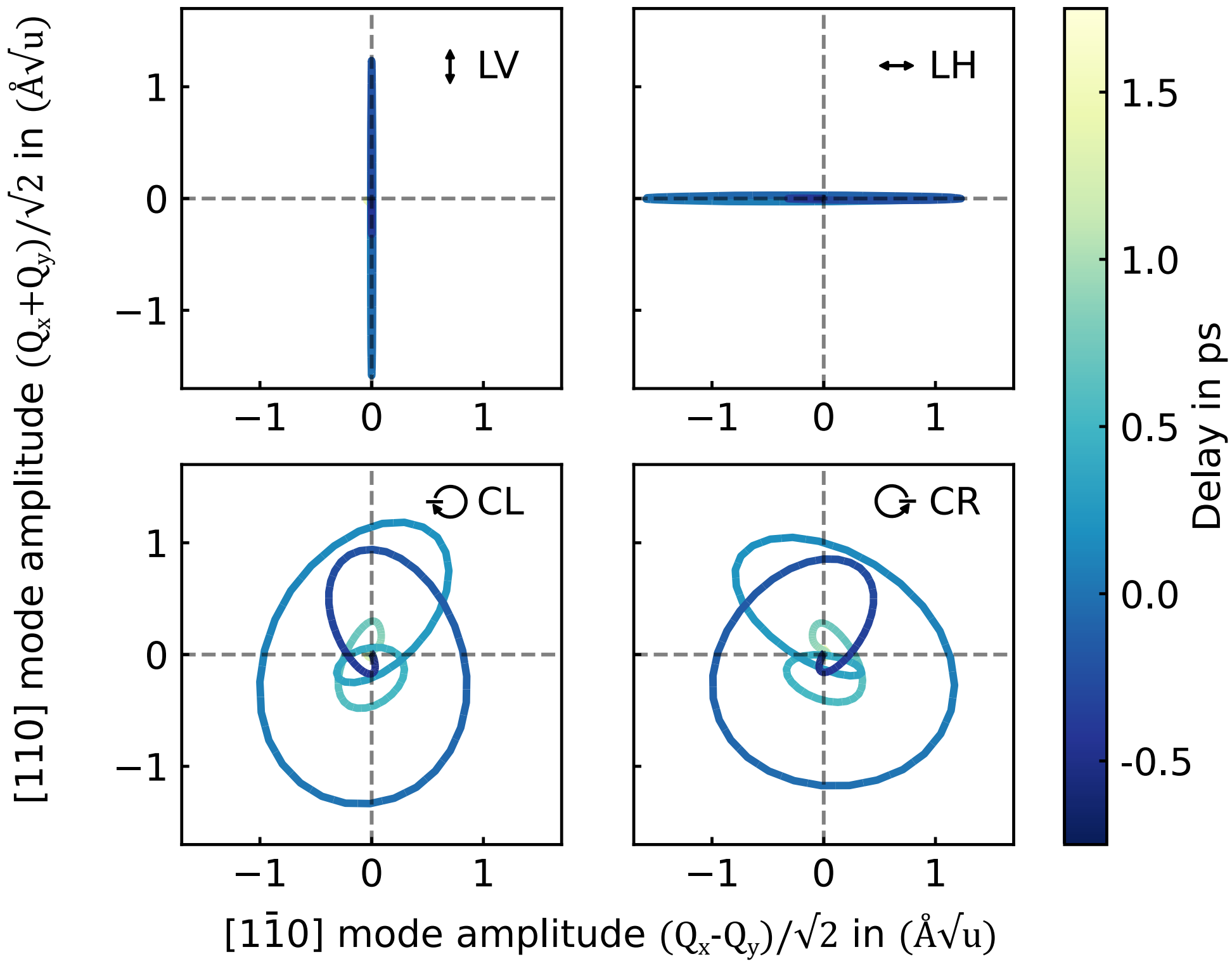

**Fig. 3: Extracted in-plane phonon mode amplitudes.** The four panels show the amplitude of the soft mode eigenvectors along the two orthogonal in-plane directions driven by THz light with linear vertical (LV), linear horizontal (LH) and circular left (CL) and right (CR) polarization from -0.5ps to 1.75ps time delay. The eigenvectors of the modes are given in Extended Data Tables 2 and 3.

The amplitudes of the in-plane eigenvectors completely describe the two-dimensional motions of the ions along the soft mode eigenvector. Fig. 3 shows the mode amplitudes over time. The eigenmode trajectories clearly follow the direction of the applied electric field for linear vertical and horizontal polarization and are circular with opposite handedness for circular left and right pump polarization. As the soft mode is strongly damped with a damping ratio of $2\Gamma/\omega \approx 1$, oscillations are only observed during the driving electric field of the THz pulse. With the temporal evolution of the mode amplitudes $Q_n(t)$, the real space time-resolved motions $\Delta\boldsymbol{r}_j$ of atom $j$ around its equilibrium position is given by $\Delta\boldsymbol{r}_j(\mathrm{t}) = Q_x(\mathrm{t})\boldsymbol{\varepsilon}_{jx} + Q_y(\mathrm{t})\boldsymbol{\varepsilon}_{jy}$. Fig. 4a shows the motions of the atoms within the STO unit cell. The oxygen and strontium ions reach amplitudes of 4 pm and 1 pm, respectively, with the same sense of rotation but opposite phases, corresponding to around 2.5% change of their bond length. Videos of these atomic motions are available in the supplementary information with an amplification of the excursion by a factor of 15 for visibility.

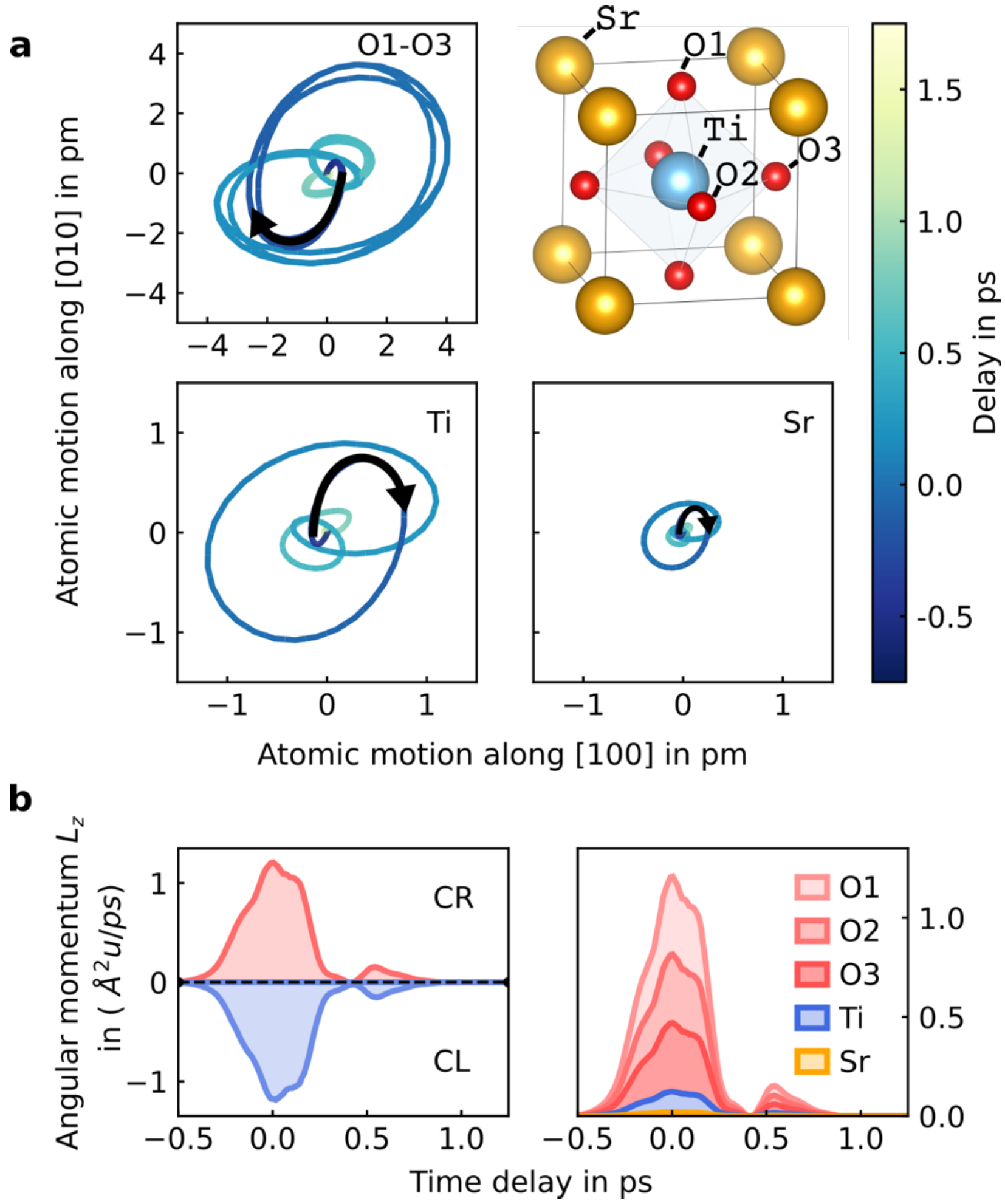

**Fig. 4: Real space atomic motions and phonon angular momentum. a)** The panels show the in-plane motions of the three negatively charged oxygen and the positively charged Sr and Ti ions in units of pm evaluated from the x-ray diffraction measurements and the eigenmode amplitude shown in Fig. 3 for circular left THz excitation. Oppositely charged ions rotate in the same circular direction (clockwise), but with opposite starting directions as indicated by the black arrows. **b)** The [001] out-of plane component of the phonon angular momentum per unit cell $L_z$ given in $Å^2u/ps$, where u is the atomic mass unit. $L_z$ changes sign with the helicity of the excitation (left panel). Individual contributions of each atom to the total phonon angular momentum shown cumulatively (right hand side).

The circular motion of the atoms induces the phonon angular momentum $\boldsymbol{L}(t)$ oriented along the out-of plane direction and defined as $\boldsymbol{L}(t) = \sum_j m_j \boldsymbol{r}_j(t) \times \dot{\boldsymbol{r}}_j(t)$. The time-dependent amplitude $L_z$ is presented in Fig. 4b. As anticipated, the sign of $L_z$ reverses for CL and CR THz polarizations. The phonon angular momentum reaches its maximum after around 250 fs as the polar mode amplitude peaks, followed by a fast decay within less than 1 ps as it is transferred to other, likely acoustic modes. Note that the time-evolution of the angular momentum can be modulated by shaping the electric field of the THz pump pulse with filters. Using a filter creating long lived oscillations of the driving THz field at 2.9 THz extends the lifetime to around 2 ps as shown in the Supplementary Materials Section 2. Separating the total angular momentum into individual atomic contributions reveals that the three oxygen atoms contribute approximately 90% of the total angular momentum despite their smaller individual masses $m_j$. This dominance arises from their larger numbers and larger trajectory radii. The resulting imbalance between the angular momentum of the negatively charged oxygens and the positively charged metal ions provides a clear explanation for both the breaking of time reversal symmetry and the net magnetic moment observed in optical experiments.

In summary, we have used femtosecond x-ray diffraction to quantify the circular phonon modes in $SrTiO_3$ in real time, reconstructing full ion trajectories with sub-picometer precision and computing the angular momentum carried by each ion. Oxygen dominates (~90%) the phonon angular momentum despite its lighter mass. These circular motions give rise to transient orbital magnetic moments associated with the lattice, providing a pathway for inducing magnetic (effective) fields in nominally non-magnetic materials. Our approach provides a general toolkit to directly quantify phonon angular momentum in solids in the time domain. Such dynamics are intimately linked to chiral phonon modes, suggesting a real-time route to probing and controlling topological phonon transport and phonon-driven magnetism in quantum materials [17,39–44].

**Acknowledgement**

We acknowledge the Paul Scherrer Institut, Villigen, Switzerland, for provision of beamtime at the Bernina beamline of SwissFEL.

RM acknowledges funding from the Swiss National Science Foundation (SNF) through the Project Funding grant no. 200021-236508.

MB acknowledges support from the Swiss National Science Foundation (SNF) through Ambitione grant no. PZ00P2_216089

J.-C.Y. acknowledges financial support from the National Science and Technology Council (NSTC), Taiwan, under Grant Nos. 114-2124-M-006-003 and 115-2923-M-006-001-MY4.

**Author contributions:**

Conceptualization: RM, SZ

Software: RM, HTL

Methodology: RM, MS, US, HTL, MF

Investigation: RM, SZ, MS, HTL, XL, DB, S-WH, MB, US

Resources: PK, J-CY

Visualization: RM

Project administration: RM

Writing – original draft: RM, US

Writing – review & editing: RM, US, MS, MB, HTL

**Competing interests:** The authors declare that they have no competing interests.

**Additional information:**

Supplementary Information is available for this paper.

All raw data files as well as scripts and derived data will be archived at discovery.psi.ch and will be published and cited here using digital object identifiers before publication. Correspondence and requests for materials should be addressed to R. Mankowsky.

# Methods & Supplementary Material

## Experimental geometry

The experimental geometry is shown in Extended Data Fig. 5. The THz pulses are generated with 1.5um, 1.1mJ, 35 fs near infrared (NIR) light pulses in an organic DSTMS crystal. Copropagating 800nm beams are separated with a dichroic mirror, which reflects the 800nm beam for active position stabilization of the laser. The linear polarization of the THz pulses (orange) can be adjusted by rotating the crystal together with the NIR beam polarization using a $\lambda/2$ waveplate (HWP). Optional band-pass or high-pass filters can be inserted into the THz beam with a filter wheel. A birefringent Y-cut quartz crystal with 89um thickness (QWP) acts as quarter waveplate at 2.9 THz converting linear to circular polarization.

The THz pulses pass through a 3mm thick Polymethylpenten (TPX) window into the vacuum chamber and are focused with 2inch focal distance parabolic mirror onto the sample. Their field strength and spectral content are characterized in vacuum by electrooptic sampling with 35fs 800nm pulses with 90µm FWHM diameter (red) in two 100um thick GaP crystals mounted next to the sample, which are oriented to measure horizontal and vertical THz polarization, respectively. The x-rays (green) are guided through a hole in the parabolic THz mirror onto the sample. The sample is mounted onto the cold finger of a Helium flow cryostat reaching sample temperatures between 5 to 450 K. The incidence angle of x-ray and THz beams is set with a rotation stage rotating the hexapod to which the cryostat is mounted. The azimuthal rotation of the sample around its surface normal is changed manually with a wobble stick.

The x-ray energy of the 50fs x-ray pulses was defined with a double crystal Si (111) monochromator to a bandwidth of ~1eV. KB mirrors were used to focus the x-rays to a spot size of 75µm diameter, around ~2 times smaller than the THz spot size of 175µm. The incidence angles of the THz pulses with respect to the sample surface and x-ray photon energies were 74.74° / 8.3keV and 102.58° / 7.5keV in diffraction condition of the $(203)$ / $(023)$ and $(223)$ peaks, respectively. These angles are close to normal incidence (90°) and the reduction of the THz electric field strength due to spot size elongation in the vertical direction is negligible. However, reflection losses are significant and must be considered when evaluating the internal field driving the mode in the STO film. Since the film thickness of 80 nm is far below the THz wavelength of 100µm at 3 THz, we are using the substrate refractive index of $n_{Si} = 3.41$ to evaluate the field transmitted into the film. Using the Fresnel coefficient at normal incidence, we obtain a value of $t = 2/(1 + n_{Si}) \approx 0.453$, reducing the peak field from 400 kV/cm to around 180 kV/cm inside the film. The vertically scattered X rays are collected with a 0.5M Jungfrau detector above the chamber. For details about the setup, see reference 33 of the main text. Circular left and circular right polarization are defined by a left- and right-handed convention around the sample surface normal.

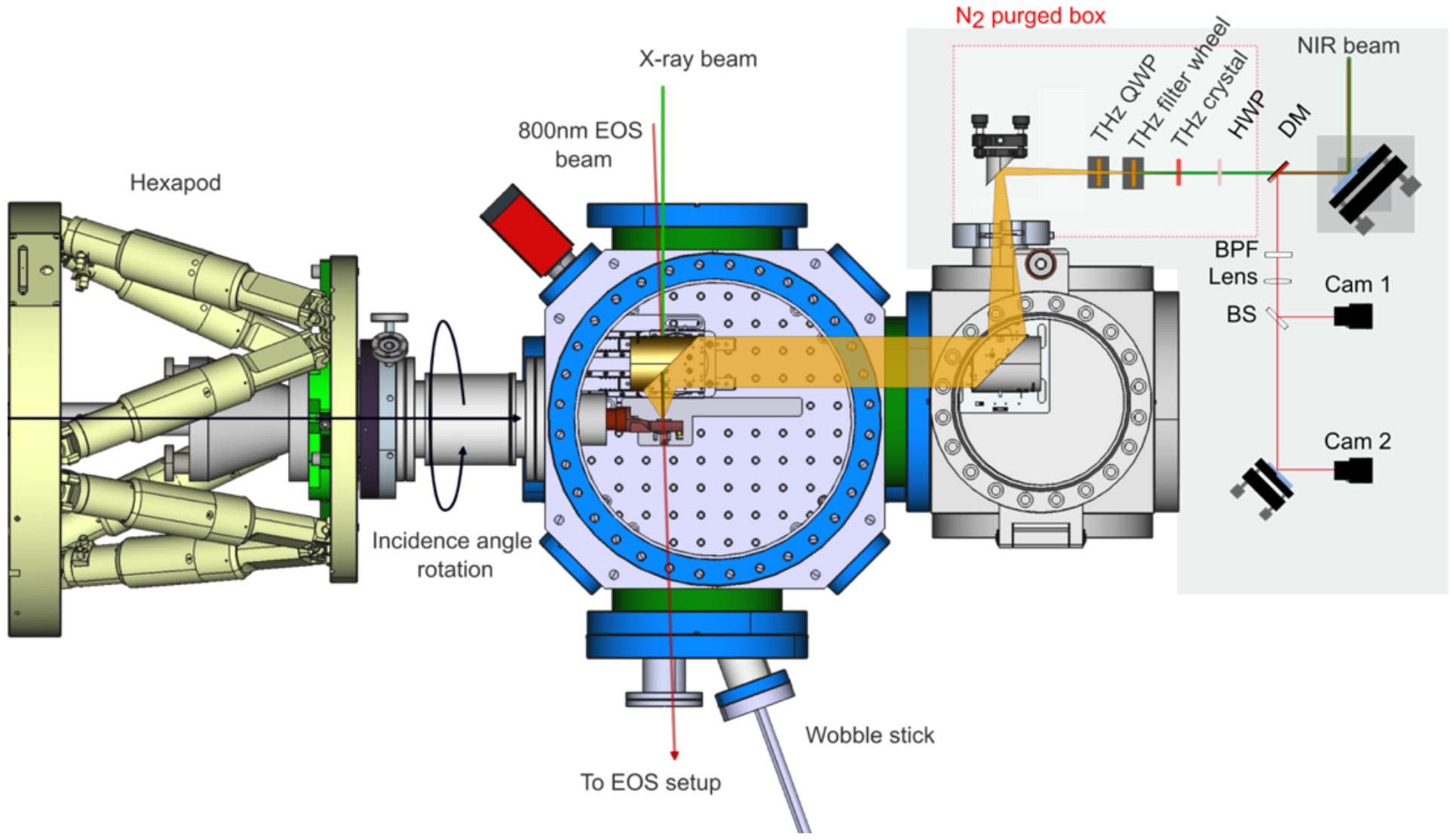

**Extended Data Fig. 5: Top view of the experimental setup.** The THz radiation (orange) is generated in a THz DSTMS crystal outside of the chamber. Their linear polarization is controlled by rotating both the crystal and the NIR beam polarization with a $\boldsymbol{\lambda/2}$ waveplate (HWP) and is converted to circular using a $\boldsymbol{\lambda/4}$ waveplate (QWP). The THz pulses (orange) are guided through a window into the vacuum chamber and are focused with a 2" focal distance parabolic mirror in vacuum onto the sample and overlapped with the probe pulses (green) on the sample.

The THz electromagnetic field was characterized by electrooptic sampling (EOS) at the sample location using 800nm pulses (red). Part of this Figure has been reproduced from reference 33 of the main text.

**THz pulse electric fields**

THz pulses were generated with 1.1mJ, 35 fs 1.5µm wavelength near-infrared (NIR) pulses in a DSTMS crystal. The linear polarization of the generated THz pulses was switched between vertical and horizontal by rotating the DSTMS crystal by 90° and adjusting the NIR polarization using a $\lambda/2$ waveplate. An additional 89µm y-cut quartz crystal was used as a $\lambda/4$ waveplate to create circular polarization at 2.9 THz.

**Characterization of the THz electromagnetic field**

To get a clean electrooptic sampling measurement, a polarizer was used to select either the vertical or horizontal component of the THz electric field. The measurement results are shown in Extended Data Fig. 6, where lines in the inset denote the orientation of the polarizer (orange) and the $\lambda/4$ waveplate (black) during the measurement.

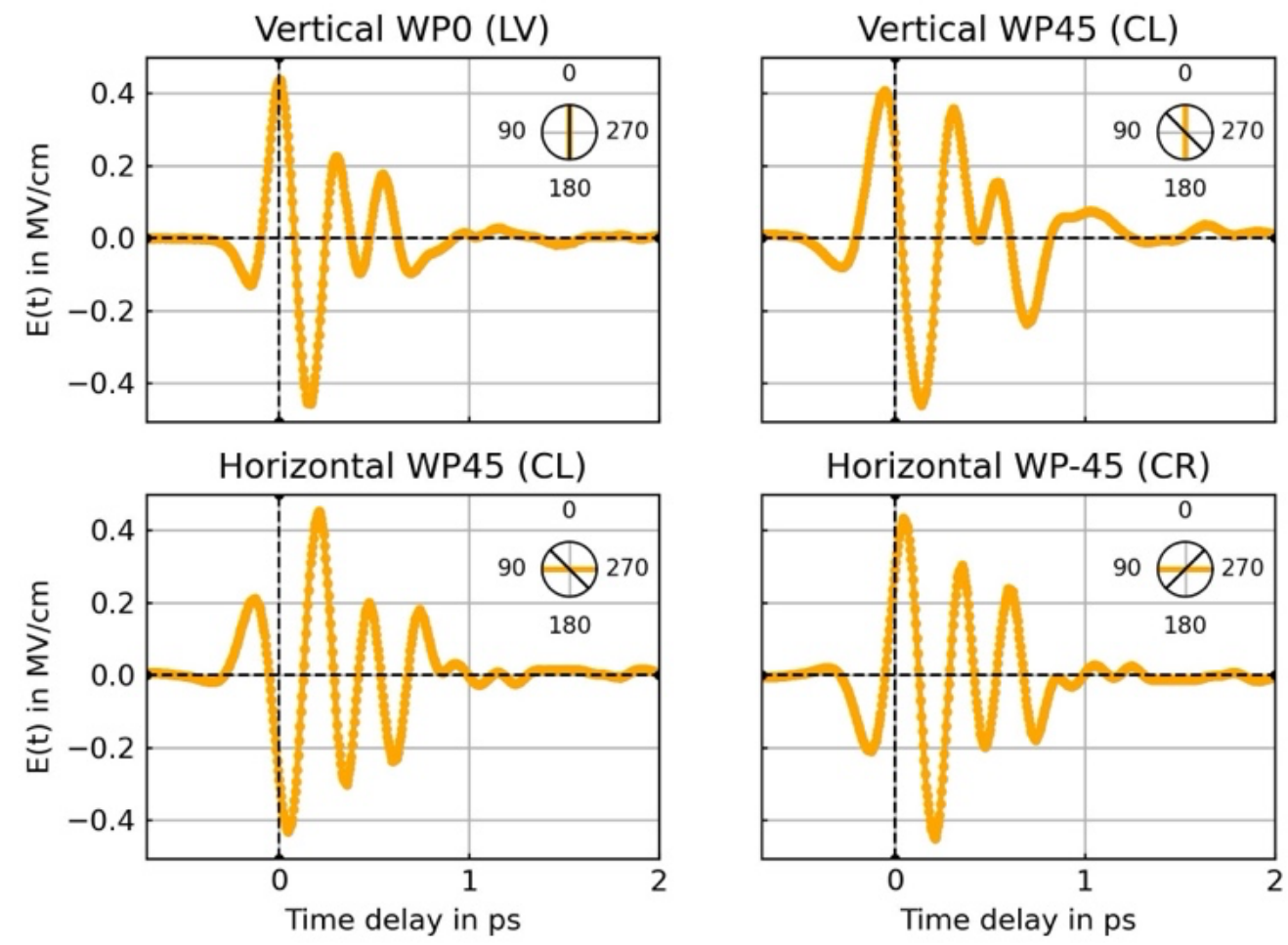

**Extended Data Fig. 6: Electric fields obtained from electrooptic sampling.** Electric field obtained from electrooptic sampling for linear vertical (LV), linear horizontal (LH), circular left (CL) and circular right (CR) THz polarization. The inset denotes the orientation of the $\boldsymbol{\lambda/4}$ waveplate (black) and the orientation of a THz polarizer inserted behind it (see text).

**Circularity of the THz field**

The horizontal projection of the measured electric field for CL polarization has higher frequencies than the vertical projection. In this measurement, the THz polarization was vertical before the $\lambda/4$ waveplate. The difference in shape and spectral content is the result of using a monochromatic waveplate, for which the retardation is only matched to create circular fields at 2.9 THz. Higher frequencies experience a larger phase change and rotate towards horizontal orientation, while lower frequencies remain less affected and are more vertically polarized. The polarization state can easily be quantified by taking the Fourier transform of the waveforms shown above.

The frequency-dependent polarization state is then simply given by:

$$P(\upsilon) = \begin{cases} 1 - 0.5\dfrac{a_\mathrm{h}}{a_\mathrm{v}}, \text{if } a_\mathrm{v} < a_\mathrm{h} \\ 0.5\dfrac{a_\mathrm{v}}{a_\mathrm{h}}, \text{if } a_\mathrm{v} > a_\mathrm{h} \end{cases}$$

$P(\upsilon)$ changes from 1 for vertical through 0.5 for circular and 0 for horizontal polarization. The two projections and the evaluated Fourier transform and frequency resolved polarization state are shown in Extended Data Fig. 7, clearly showing the different spectral content and the changing polarization state across the spectrum.

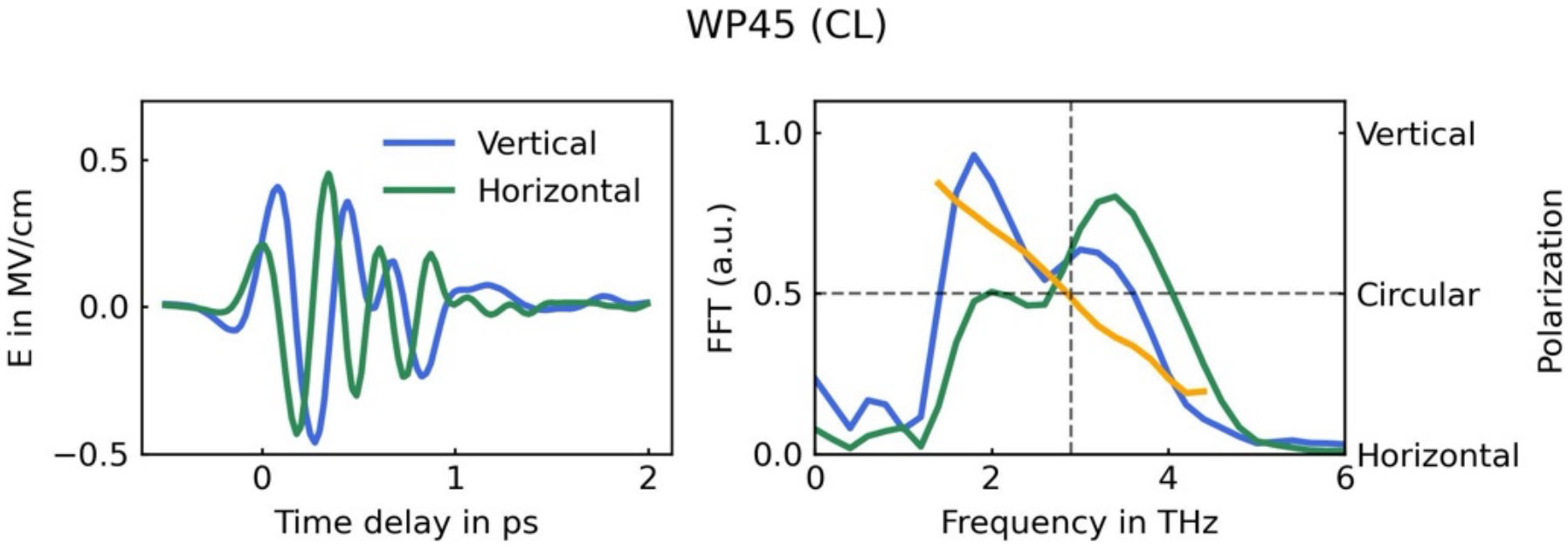

**Extended Data Fig. 7: Polarization states of the electromagnetic field.** Projections of the electric field onto the vertical and horizontal direction for CL polarization created from vertical polarization with a monochromatic $\boldsymbol{\lambda/4}$ waveplate. The right hand side shows the Fourier transform amplitude as well as the derived frequency dependent polarization state $\boldsymbol{P(\upsilon)}$ (see text for details).

**Characterization of the THz electromagnetic field with 3 THz low pass filter**

The shape and spectral content of the THz field can be shaped with filters. In the following, we show the same characterization as above when using a 3 THz low-pass filter used for the experiments shown in section 21. The cut-off creates long lived oscillations at around 3THz, which results in a smaller bandwidth and a cleaner circular polarization state, similar to the setup used in reference [9] of the main text.

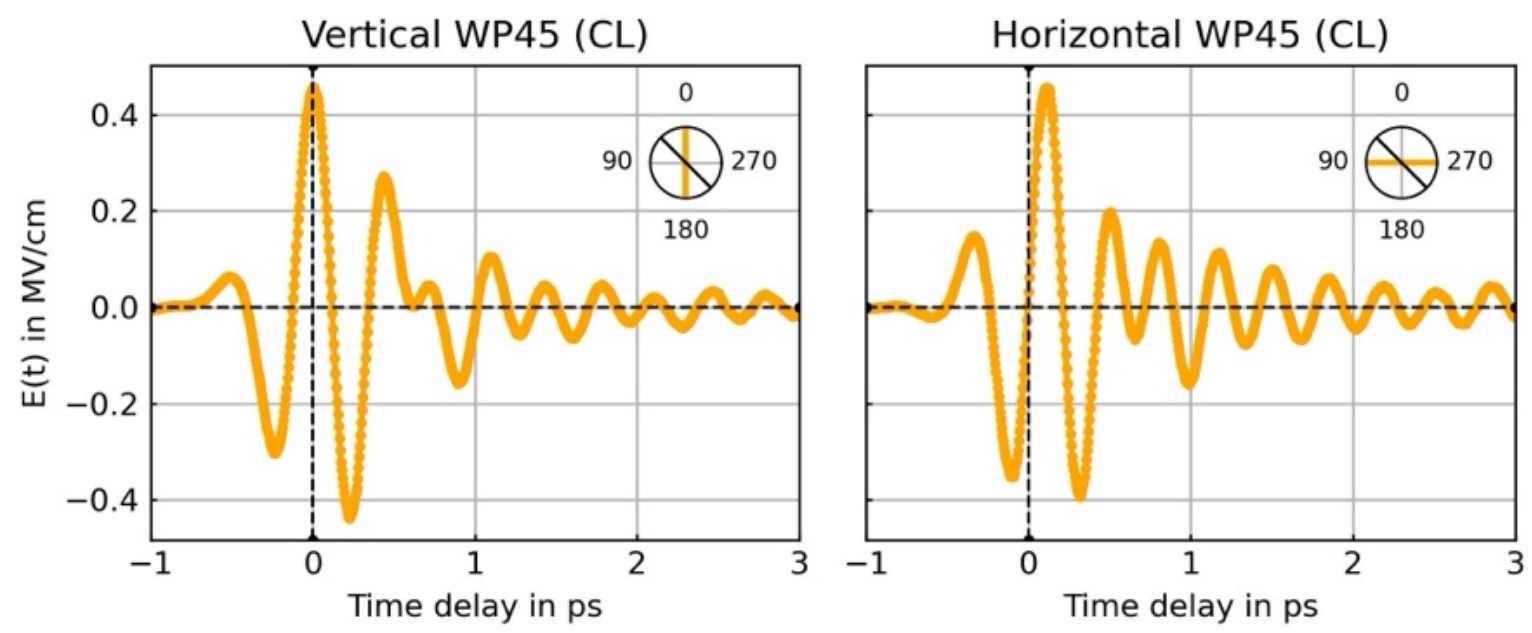

**Extended Data Fig. 8: Electric fields obtained from electrooptic sampling with low-pass filter.** Electric field obtained from electrooptic sampling for linear vertical (CL) THz polarization. The inset denotes the orientation of the $\boldsymbol{\lambda/4}$ waveplate (black) and the orientation of a THz polarizer inserted behind it (see text).

**Circularity of the THz field with 3 THz low pass filter**

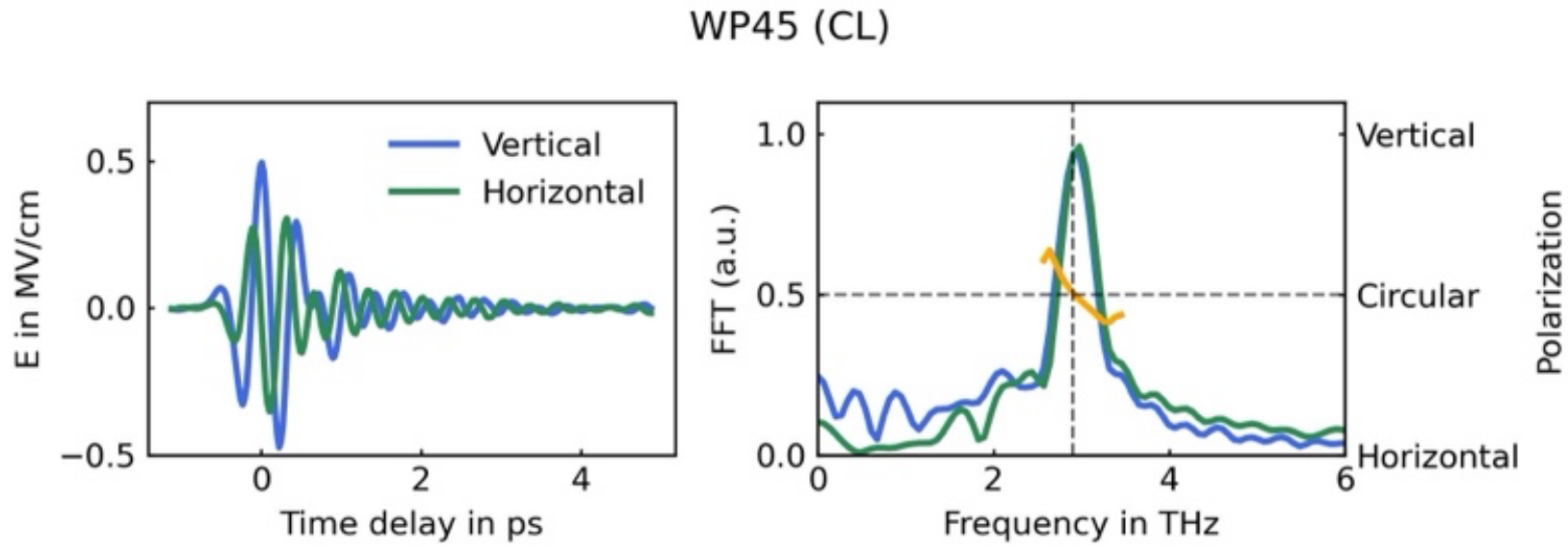

**Extended Data Fig. 9: Polarization states of the electromagnetic field with low-pass filter.** Projections of the electric field onto the vertical and horizontal direction for CL polarization created from vertical polarization with a monochromatic $\boldsymbol{\lambda/4}$ waveplate. The right hand side shows the Fourier transform amplitude as well as the derived frequency dependent polarization state $\boldsymbol{P(\upsilon)}$ (see text for details).

**Data analysis**

**First-principal calculations of phonon eigenvectors of SrTiO3**

First-principles calculations were performed within the framework of density functional theory (DFT) to investigate the zone-center phonon eigenvectors of $SrTiO_3$. All calculations were carried out using the Vienna *ab initio* Simulation Package (VASP, version 6.0)[1–3] in combination with the Phonopy software package [4].

The electron–ion interaction was described using pseudopotentials constructed within the projected augmented wave (PAW) method[5]. The valence configurations included Sr $4s^2$ $4p^6$ $5s^2$, Ti $3s^2$ $3p^6$ $3d^2$ $4s^2$, and O $2s^2$ $2p^4$. Exchange–correlation effects were treated using the revised Perdew–Burke–Ernzerhof generalized gradient approximation (PBEsol) functional[6].

After convergence testing, Brillouin-zone integrations were performed using a 12x12x12 Monkhorst–Pack k-point mesh [7] together with a plane-wave energy cutoff of 600 eV. Self-consistent field calculations were iterated until the total energy converged to within $10^{-8}$ eV. Zone-center phonon eigenvectors were calculated using a 2x2x2 supercell of cubic $SrTiO_3$. All calculations were verified to be well converged with respect to the numerical parameters.

**Structure and atomic positions of $SrTiO_3$**

In the paraelectric phase, $SrTiO_3$ crystallizes in a cubic perovskite structure. The equilibrium atomic positions are given in Extended Data Table 1 in fractional coordinates. The real space positions $\boldsymbol{r}_{j0}$ can be calculated by multiplying the tabulated values with the lattice constant of $a = 3.895$Å.

| | a | b | c |
|---|---|---|---|
| **Sr** | 0.0 | 0.0 | 0.0 |
| **Ti** | 0.5 | 0.5 | 0.5 |
| **O1** | 0.5 | 0.5 | 0.0 |
| **O2** | 0.5 | 0.0 | 0.5 |
| **O3** | 0.0 | 0.5 | 0.5 |

**Extended Data Table 1: equilibrium atomic positions in fractional coordinates.**

**Eigenvectors: Phonon mode displacements**

In the notation below, we are using the indices $\boldsymbol{n}$ to indicate the direction of the eigenvectors: $x = [\mathbf{100}]$, $y = [\mathbf{010}]$, $xy = [\mathbf{110}]$ and $-xy = [\bar{\mathbf{1}}\mathbf{10}]$. Atomic displacements are given by the product of the phono mode eigenvector $\tilde{\boldsymbol{\varepsilon}}_{\boldsymbol{n}}$ listed in Extended Data Table 2 and its mass-weighted normal mode amplitude $Q_n$ in units of Å$\sqrt{u}$, where $u$ denotes the atomic mass unit. The eigenvectors are normalized such that $\sum_j \tilde{\boldsymbol{\varepsilon}}_j^2 m_j = 1u$, where $m_j$ is the *j*th atoms mass in units of $u$. The atomic displacements $\boldsymbol{r}_{nj}$ of the *j*th atom for an amplitude $Q_n$ of the phonon mode with index *n* in units of Å are given by: $\boldsymbol{r}_{nj} = Q_n \cdot \tilde{\boldsymbol{\varepsilon}}_{\boldsymbol{nj}}/\sqrt{m_j}$. The eigenvectors $\boldsymbol{\varepsilon}$ listed in Extended Data Table 3 include this correction, simplifying the relation to: $\boldsymbol{r}_{nj} = Q_n \cdot \boldsymbol{\varepsilon}_{nj}$.

| | $\tilde{\boldsymbol{\varepsilon}}_x$ $[\mathbf{100}]$ | | | $\tilde{\boldsymbol{\varepsilon}}_y$ $[\mathbf{010}]$ | | | $\tilde{\boldsymbol{\varepsilon}}_{xy}$ $[\mathbf{110}]$ | | | $\tilde{\boldsymbol{\varepsilon}}_{-xy}$ $[\bar{\mathbf{1}}\mathbf{10}]$ | | |
|---|---|---|---|---|---|---|---|---|---|---|---|---|
| **Sr** | 0.03 | 0.00 | 0.00 | 0.00 | 0.03 | 0.00 | 0.02 | 0.02 | 0.00 | -0.02 | 0.02 | 0.00 |
| **Ti** | 0.07 | 0.00 | 0.00 | 0.00 | 0.07 | 0.00 | 0.05 | 0.05 | 0.00 | -0.05 | -0.05 | 0.00 |
| **O1** | -0.13 | 0.00 | 0.00 | 0.00 | -0.13 | 0.00 | -0.09 | -0.09 | 0.00 | 0.09 | 0.09 | 0.00 |
| **O2** | -0.13 | 0.00 | 0.00 | 0.00 | -0.11 | 0.00 | -0.09 | -0.08 | 0.00 | 0.09 | 0.08 | 0.00 |
| **O3** | -0.11 | 0.00 | 0.00 | 0.00 | -0.13 | 0.00 | -0.08 | -0.09 | 0.00 | 0.08 | 0.09 | 0.00 |

**Extended Data Table 2: Polar mode eigenvectors as obtained from DFT calculations.** The dimensionless eigenvectors above describe the motions of the polar mode along the in-plane directions $\boldsymbol{n} = [\mathbf{100}]$, $[\mathbf{010}]$, $[\mathbf{110}]$ and $[\bar{\mathbf{1}}\mathbf{10}]$ and are normalized such that $\sum_j \tilde{\boldsymbol{\varepsilon}}_j^2 \boldsymbol{m}_j = \mathbf{1}\boldsymbol{u}$, where $\boldsymbol{u}$ is the atomic mass unit and $\boldsymbol{m}_j$ the mass of the *j*th atom in $\boldsymbol{u}$. Note that due to the normalization, each atomic displacement vector must be divided by the respective atomic mass to obtain the actual displacements $\boldsymbol{r}_{nj}$ for an amplitude $\boldsymbol{Q}_n$: $\boldsymbol{r}_{nj} = \boldsymbol{Q}_n \cdot \tilde{\boldsymbol{\varepsilon}}_{nj}/\sqrt{\boldsymbol{m}_j}$ in Å (see text).

| | $\varepsilon_x$ [100] | | | $\varepsilon_y$ [010] | | | $\varepsilon_{xy}$ [110] | | | $\varepsilon_{-xy}$ [$\bar{1}$10] | | |
|---|---|---|---|---|---|---|---|---|---|---|---|---|
| **Sr** | 0.003 | 0.000 | 0.000 | 0.000 | 0.003 | 0.000 | 0.002 | 0.002 | 0.000 | -0.002 | 0.002 | 0.000 |
| **Ti** | 0.009 | 0.000 | 0.000 | 0.000 | 0.009 | 0.000 | 0.007 | 0.007 | 0.000 | -0.007 | 0.007 | 0.000 |
| **O1** | -0.032 | 0.000 | 0.000 | 0.000 | -0.032 | 0.000 | -0.023 | -0.023 | 0.000 | 0.023 | -0.023 | 0.000 |
| **O2** | -0.032 | 0.000 | 0.000 | 0.000 | -0.028 | 0.000 | -0.023 | -0.020 | 0.000 | 0.023 | -0.020 | 0.000 |
| **O3** | -0.028 | 0.000 | 0.000 | 0.000 | -0.032 | 0.000 | -0.020 | -0.023 | 0.000 | 0.020 | -0.023 | 0.000 |

**Extended Data Table 3: Polar mode eigenvectors corrected by the atomic masses.** These eigenvectors are obtained from the DFT calculation results of Extended Data **Table 2** and have a unit of $1/\sqrt{\boldsymbol{u}}$, yielding atomic displacements in Å when multiplied with the mass-weighted normal coordinate $Q$.

Note that there is a simple relation between the eigenvectors along the diagonal and the principal axis: $\boldsymbol{\varepsilon}_{[110]} = (\boldsymbol{\varepsilon}_{[100]} + \boldsymbol{\varepsilon}_{[010]})/\sqrt{2}$ and $\boldsymbol{\varepsilon}_{[1\bar{1}0]} = (\boldsymbol{\varepsilon}_{[100]} - \boldsymbol{\varepsilon}_{[010]})/\sqrt{2}$

**Structure factor calculations**

To obtain the polar mode amplitudes from the measured diffraction intensity $I_{hkl} \propto |F_{hkl}|^2$, we calculate the respective modulation of the diffraction peak structure factors $F_{hkl} = \sum_j f_j e^{-i\boldsymbol{G}_{hkl}\boldsymbol{r}_j}$. Here, $f_j$ and $\boldsymbol{r}_j(\mathrm{t}) = \boldsymbol{r}_{j0} + Q_n(\mathrm{t})\boldsymbol{\varepsilon}_{nj}$ are the scattering factor and fractional position of the $j$-th atom in the unit cell, and $\boldsymbol{G}_{hkl}$ is the reciprocal lattice vector of the diffraction peak with miller indices $hkl$.

The ratio of the diffracted intensity as function of the mode amplitude $Q$ can be calculated as follows and is shown in Extended Data Fig. 10 for the measured (203) / (023) and (223) peaks and eigenvectors with in-plane displacements along [100], [010] and [110], [$\bar{1}$10], respectively:

$$\frac{I}{I_0}(Q) = \frac{|F_{hkl}(Q)|^2}{|F_{hkl}(Q=0)|^2}$$

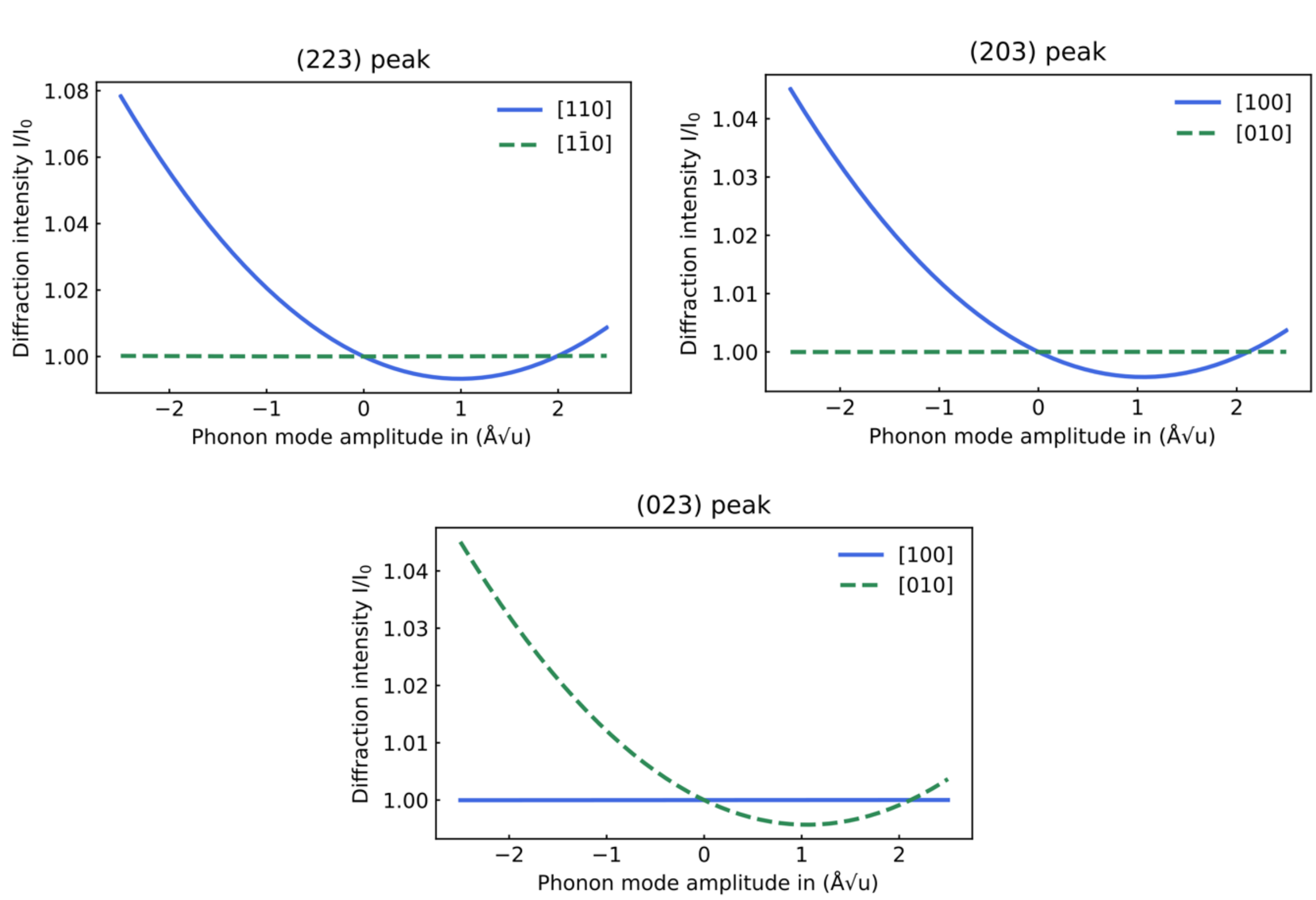

**Extended Data Fig. 10. Structure factor calculations.** Calculated diffracted intensity for varying amplitude of the degenerate ferroelectric soft mode excited along the in-plane $[\mathbf{110}]$, $[\bar{\mathbf{1}}\mathbf{10}]$ and $[\mathbf{100}]$, $[\mathbf{010}]$ directions for the measured $(\mathbf{223})$, $(\mathbf{203})$ and $(\mathbf{023})$ diffraction peaks, respectively. The eigenvectors used for these calculations are given in Extended Data **Table 3**.

**Background subtraction**
The background signal observed on the peaks consists of two components captured in the function $g(t)$ below. The first describes a rise and decay, setting in shortly after the main part of the THz electric field. The second component is the longitudinal strain wave, which forms at the STO / air interface as the polar mode decays and the energy is transferred into acoustic modes:

$$g(t) = \frac{A_0}{2} \cdot \left(1 + erf\left((t - t_0)/dt\right)\right) \cdot \exp\left(-\frac{(t-t_0)}{\tau}\right) + \frac{A_1}{2} \cdot (1 + erf(t - t_1)/\tau) \cdot \left(1 - cos \cdot \left(\omega(t - t_1)\right)\right)$$

Here, $A_i$, $t_i$ and describe the amplitude and onset time of the signals, $dt$ and $\tau$ the risetime and lifetime of the first component and $\omega$ a frequency describing the strain wave component. The fitted background is shown in Extended Data Fig. 11 as dashed line together with the experimental data of the intensity change of the $(2\bar{2}3)$ peak for excitation with circular right THz polarization at early times up to 3ps and for longer delays up to 8ps, where the intensity change due to the onset of the strain induced peak shift becomes the dominant signal.

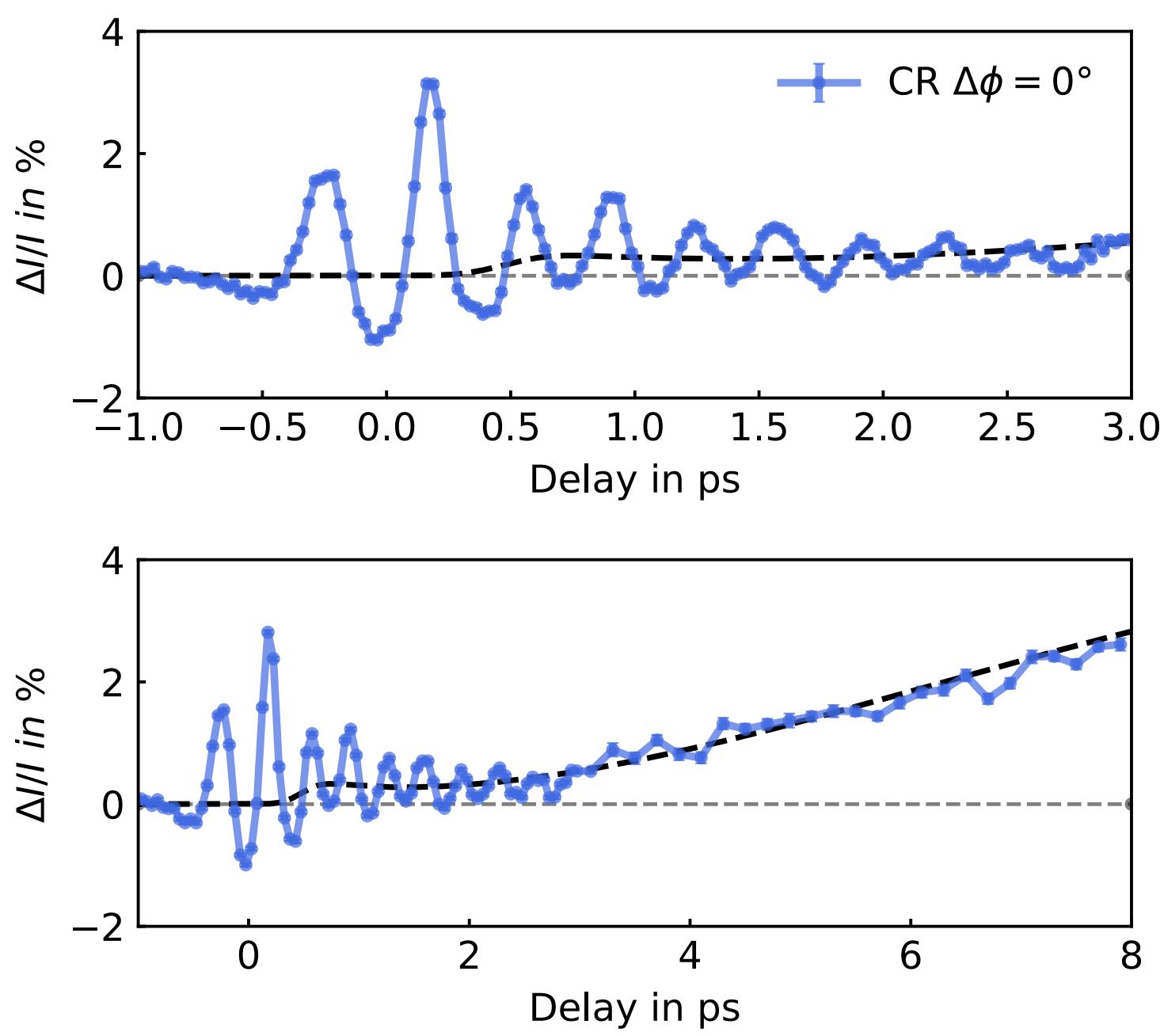

**Extended Data Fig. 11: Background function.** Intensity changes of the $(\mathbf{2\bar{2}3})$ peak for excitation with circular right THz polarization (blue markers) and the fitted background function (dashed line) following a functional form described in the text.

**Methods References**

## Supplementary Information

**THz – Time-Domain Spectroscopy**

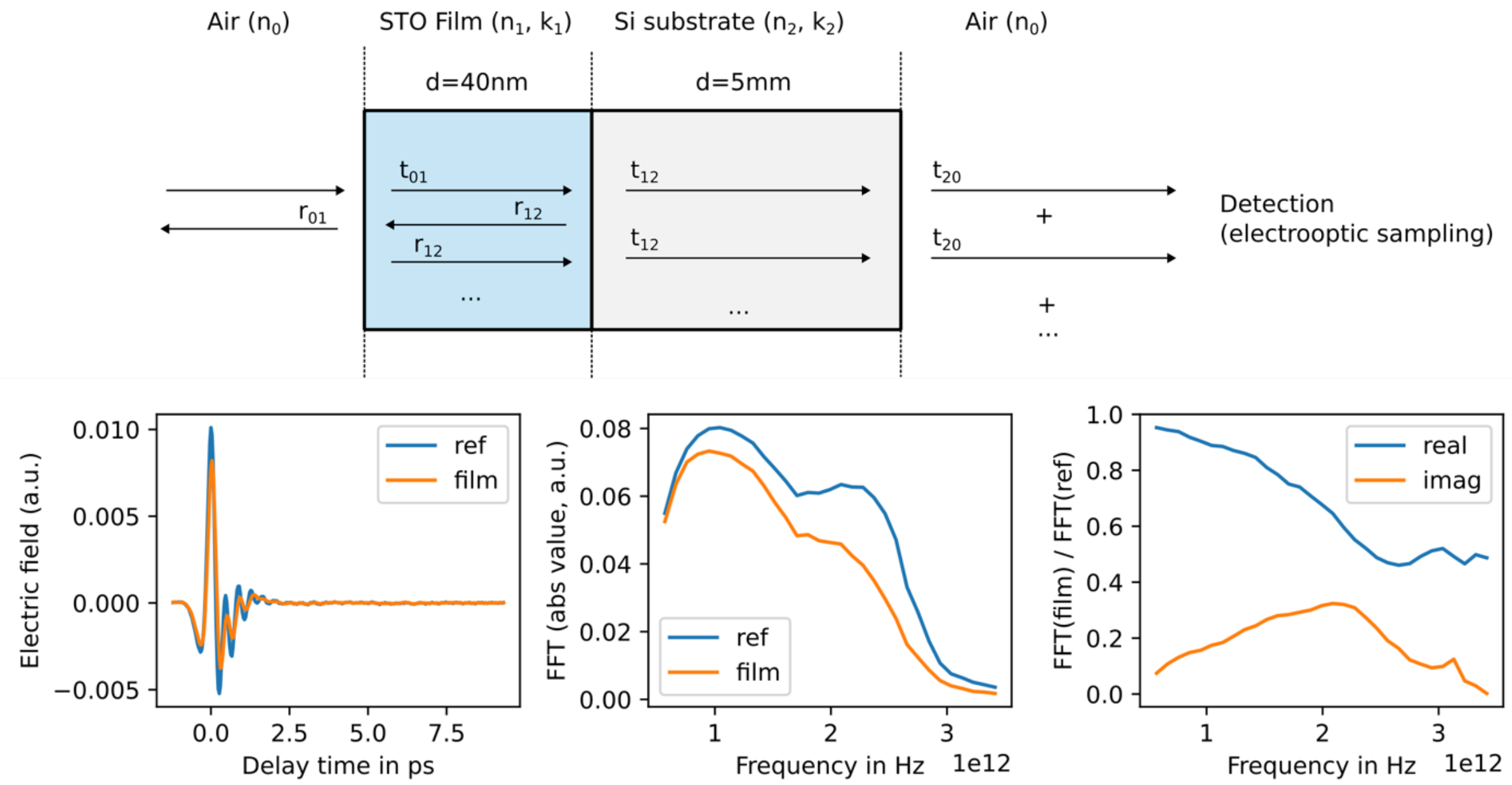

**Supplementary Fig. 1: THz time domain measurements.** The THz field as measured by EOS in transmission once through the 40nm STO thin film on a 5mm Si substrate (labeled film) and through only the substrate (labeled ref). The lower panels show the recorded THz transients (left), the absolute value of their Fourier Transform (middle) and the complex ratio of the Fourier Transforms of the measurements with and without film.

We extract the optical properties of the soft mode oscillator from the raw THz waveforms provided by the authors of reference [32], measured on a 40nm freestanding $SrTiO_3$ film. In ref [32], the THz electric field was measured twice by EOS: first, in transmission through the film and the 5mm Silicon substrate (denoted with the index film) and separately only through the Si Substrate (denoted ref). Both EOS measurements, their Fourier transform as well as the complex ratio $R = T_{film}/T_{ref}$ of their Fourier transform are shown in the lower panels of Supplementary Fig. 1. Thus, the interface between Air/Film, Film/Silicon and Silicon/Air have to be taken into account when modelling the experimental data. The only unknown quantity is the complex refractive index $\tilde{n}_1 = n_1 + ik_1$ of the film. Taking multiple reflections in the film into account and using the Fresnel equations at normal incidence, the ratio of the transmitted field through the film normalized to the reference can be written as:

$$R = \frac{T_{film}}{T_{ref}} = \frac{t_{01}P_1t_{12}FP}{t_{02}},$$

where $t_{ij}$, $r_{ij}$, $P_i$ and $FP$ denote the Fresnel field transmission and reflection coefficients, the propagation factor and the Fabry Perot factor, given by:

$$t_{ij} = 2\tilde{n}_i/(\tilde{n}_i + \tilde{n}_j)$$

$$r_{ij} = (\tilde{n}_i - \tilde{n}_j)/(\tilde{n}_i + \tilde{n}_j)$$

$$P_i = \exp(i\omega/c \cdot \tilde{n}_i d)$$

$$FP = 1/(1 - r_{12}P_1^2 r_{10})$$

Note that the refractive index of the film is frequency-dependent $\tilde{n}_1(\omega)$, while the refractive index of Silicon is constant across the THz spectrum. We describe the permittivity of the $SrTiO_3$ film using a Lorentz model for the soft mode of $SrTiO_3$ as commonly used in time domain THz spectroscopy measurements of $SrTiO_3$ and similar materials:

$$\varepsilon(\omega) = \varepsilon_\infty + \frac{S \cdot \omega_0^2}{(\omega_0^2 - \omega^2 - i\Gamma)}$$

The refractive index is simply the square root of the permittivity $\tilde{n}_1(\omega) = \sqrt{\varepsilon(\omega)}$. The fit and the experimental data are shown together with the residual of the real and imaginary parts in the upper panels of Supplementary Fig. 2. The remaining panels show the permittivity $\varepsilon$, refractive index $\tilde{n}_1$, the intensity reflectivity for an 80 nm film, the substrate and an infinitely thick film (i.e. an STO single crystal), and the attenuation length using $f = \omega/2\pi$ as $x$-axis and calculated using the fit results:

$$\varepsilon_\infty = 5.21$$
$$S = 1121$$
$$\omega_0/2\pi = 3.35\ \text{THz}$$
$$\Gamma/2\pi = 1.91\ \text{THz}$$

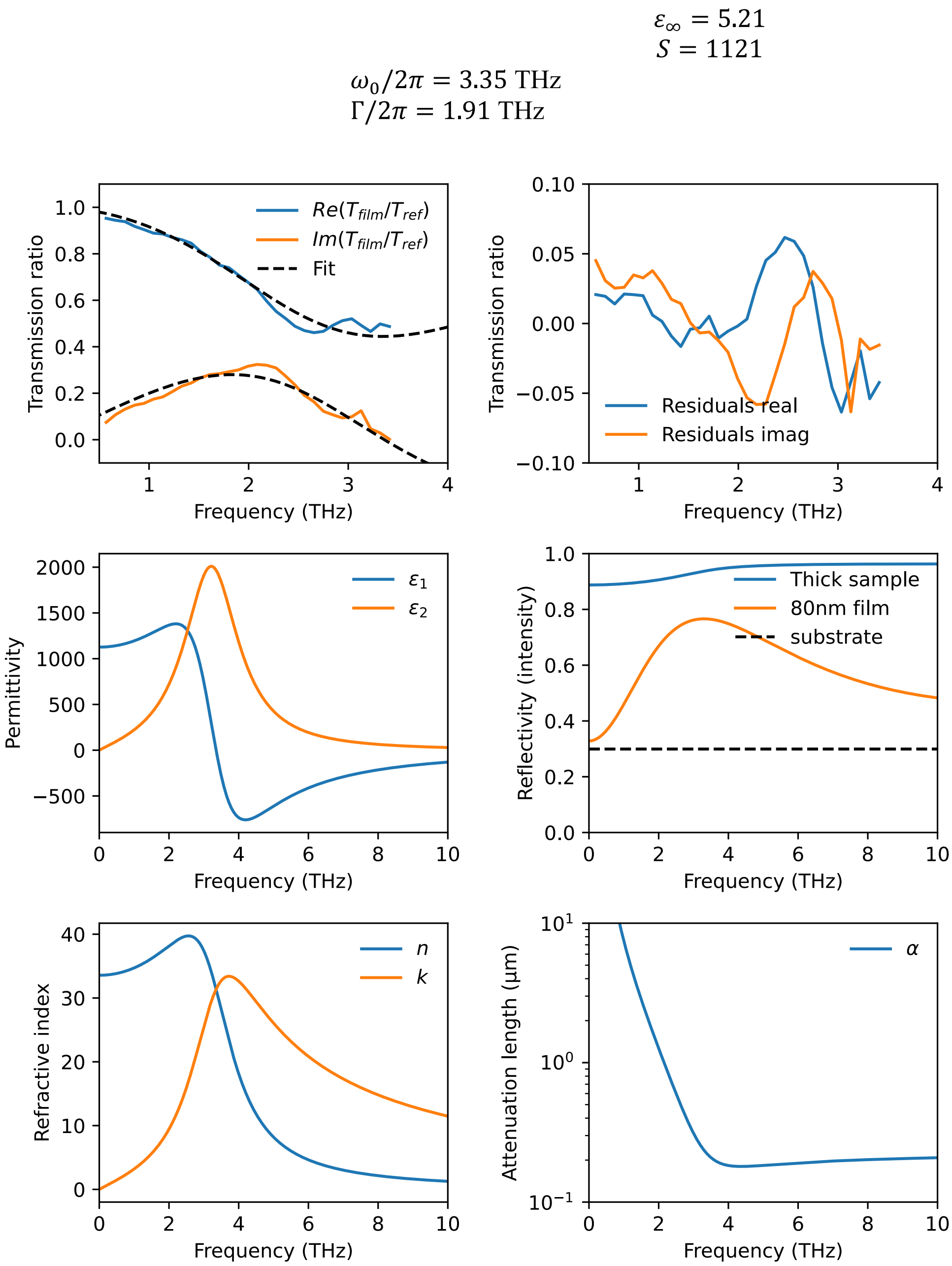

**Supplementary Fig. 2: Fit results. a, b**, Fit to the complex ratio of the Fourier Transforms of the measured THz waveforms with and without film and the residuals. **b-e**, Calculated complex permittivity, reflectivity of the thin film, substrate and an infinitely thick sample, complex refractive index modelled for the substrate, and the attenuation length.

**Circular motions and angular momentum with LP filter**

As demonstrated in the main text, the atomic motions follow the driving THz field since the polar mode of $SrTiO_3$ is strongly damped. This can be clearly demonstrated by modulating the THz exciting field with a filter. In our experiment, we use a commercial 3THz lowpass filter, which generates long-lived oscillations of the THz electric field at 2.9 THz due to the sharp frequency cut-off as shown in section 15.

Considering that the simultaneous measurement of two diffraction peaks is not always possible, we demonstrate an alternative technique to quantify the two-dimensional atomic motions, which requires only the measurement of a single diffraction peak at a time. By changing the photon energy to 7.522keV, the $(223)$ diffraction peak can be measured with a THz incidence angle of 102.58°, closer to normal incidence. To probe the atomic motions along two orthogonal directions, we first measure the $(223)$ peak, which is sensitive to in-plane motions along the $[110]$ direction, and advance the carrier envelope phase of the circular THz field by $\pi/2$ as indicated in the label of Figure 3. Since $SrTiO_3$ is centrosymmetric, the resulting measurement describes dynamics of the atoms along the orthogonal $[\bar{1}10]$ (-xy) direction.

As shown in Supplementary Fig. 3, we observe a phase change of the dynamics of $\pm\pi/2$ for circular right and left polarization, respectively, reminiscent of the dynamics shown in Figure 2 of the main text. The long-lived oscillations observed in the diffraction intensity change are a direct consequence of the longer driving electric field of the pump pulse. Using the same procedure as described in the main text, we fit the data (solid line) to obtain the amplitudes of the eigenmodes along the$[110]$ and $[\bar{1}10]$ directions. The two lower panels show the two-dimensional atomic motions for CL and CR polarized excitation, which now follow a more circular trajectory with the same amplitude as observed without the low-pass filter.

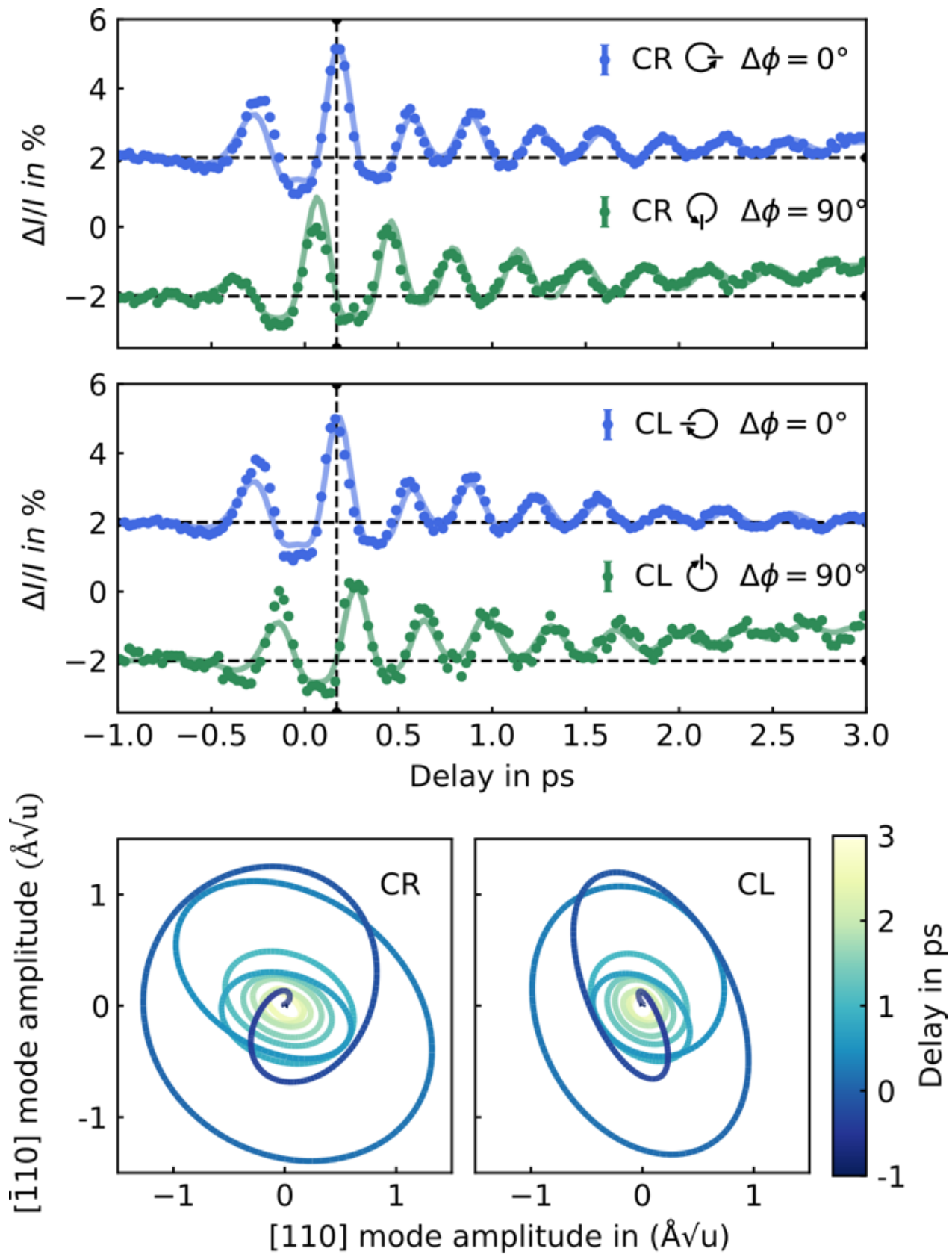

**Supplementary Fig. 3: THz polarization dependence of the diffraction intensity and extracted in-plane atomic trajectories for long driving fields.** Time-resolved changes in diffracted intensity of the $(\mathbf{223})$ peak, showing $\pm\boldsymbol{\pi}/\mathbf{2}$ phase shifts when advancing the THz carrier envelope phase by $\Delta\Phi = 90°$ for circular right (CR) and left (CL) driven atomic motions. The solid blue and green lines show the calculated change in diffraction intensity obtained from a simulation of the atomic motions using the THz field obtained from electrooptic sampling. The traces are vertically offset by $\pm\mathbf{2}\%$ for better visibility. The lower panels show the amplitudes of eigenmodes $\boldsymbol{Q}_{xy}$ and $\boldsymbol{Q}_{-xy}$ of the polar soft mode along the two orthogonal in-plane directions.

Equivalent to the procedure described in the main text, we derive the atomic motions from the amplitudes using the eigenvectors given in Extended Data Table 3. The oxygen and strontium ions reach amplitudes of 4 pm and 1 pm, respectively, with the same sense of rotation but opposite phases, corresponding to around a 2.5% change of their bond length. However, their motion quickly decays to long lived oscillations, which are about 1/3 in amplitude as shown in **Supplementary Fig. 4**. The circular motions induce an angular momentum for a longer time as shown in the lower panels of the same figure, caused by the longer driving pulse. Note that the decay time itself remains the same.

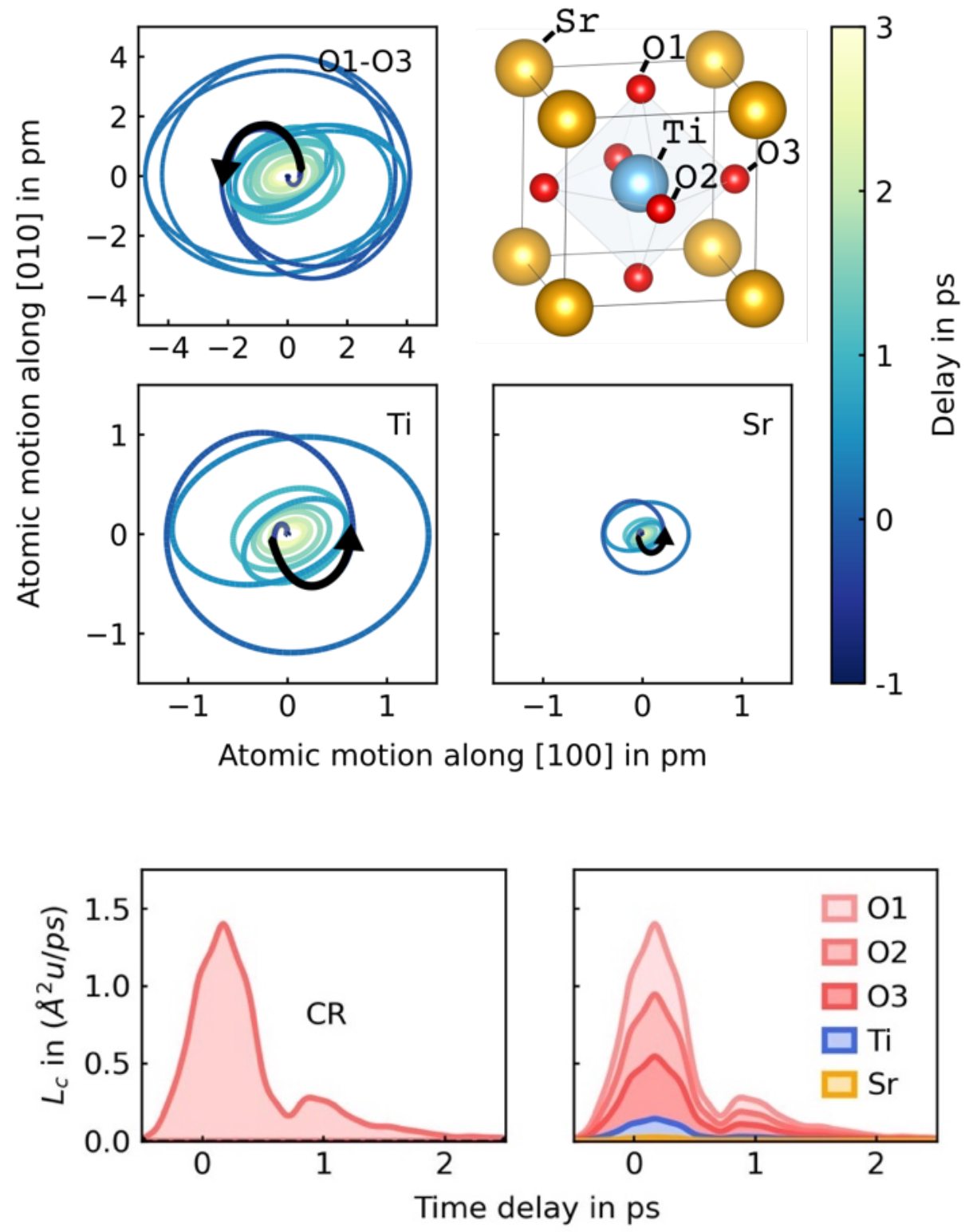

**Supplementary Fig. 4: Real space atomic motions and phonon angular momentum.** The upper panels show the in-plane motions of the three negatively charged oxygen and the positively charged Sr and Ti ions in units of pm extracted from the x-ray diffraction measurements with circular right polarized THz excitation of Supplementary Fig. 3. Note that oppositely charged ions move with the same sense of rotation but opposite starting directions as indicated by black arrows. The lower two panels show the [001] out-of plane component of the phonon angular momentum per unit cell $\boldsymbol{L_c}$ extracted from the motions shown in the panels above. The right hand side illustrates the individual contributions of each atom to the total phonon angular momentum cumulatively (right hand side).

Finally, we investigate any signatures of circular dichroism signal on top of the structural dynamics due to the soft mode motion. Such signals would occur along the out-of plane direction of the sample and appear in all the measured peaks since they have an L component in their miller indices. For these measurements, the diffraction geometry was adjusted such that the in-plane projection of the scattering vector is vertical, parallel to the linear THz polarization before the $\lambda/4$ waveplate. Under this condition, the oscillatory signal of the polar mode is independent of the sense of direction (CL or CR), and does not appear in the difference of the two traces as only the component perpendicular to the scattering vector changes phase. The signals as well as their differences (orange) of the measured intensity change induced by CL (blue) and CR (green) THz excitation are shown in **Supplementary Fig. 5** for the (223) and (023) peaks. Within our accuracy, no circular dichroism signals are found, which is to be expected considering that the measurement is insensitive to (effective) magnetic fields inside the sample and would only appear if structural rearrangements occur along the L direction due to the circular atomic motions.

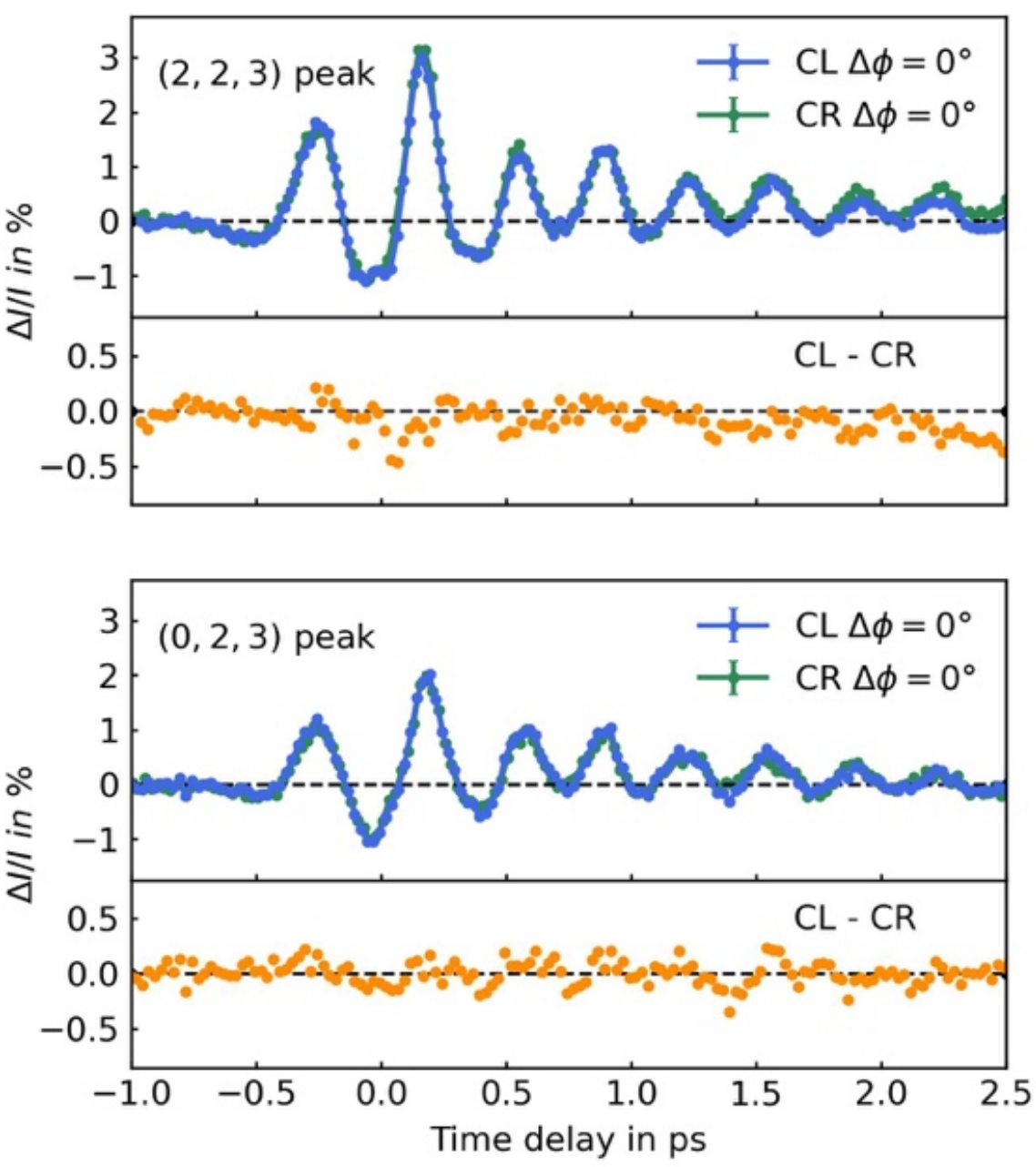

**Supplementary Fig. 5: Circular dichroism.** Change in diffraction peak intensity of the $(\mathbf{223})$ and $(\mathbf{023})$ peaks for excitation with THz pulses of CL (blue) and CR (green) polarization and their difference (orange).